\documentclass{article} 
\usepackage{nips15submit_e,times}
\usepackage{hyperref}
\usepackage{url}
\usepackage{graphicx}
\usepackage{lipsum}

\title{Multi Level Dense Layer Neural Network Model for Housing Price Prediction}

\author{
Robert Wijaya\\
College of Engineering and Computer Science\\
The Australian National University\\
Canberra, ACT 2601 \\
\texttt{Robert.Wijaya@anu.edu.au} \\
}

\nipsfinalcopy 

\begin{document}

\maketitle

\begin{abstract}
Predicting the price of a house remains a challenging issue that needs to be addressed. Research has attempted to establish a model with different methods and algorithms to predict the housing price, from the traditional hedonic model to a neural network algorithm. However, many existing algorithms in the literature are proposed without any fine-tuning and customization in the model. In this paper, the author attempted to propose a novel neural network-based model to improve the performance of housing price prediction. Inspired by the modular neural network, the proposed model consists of a three-level neural network that is capable to process information in parallel. The author compared several state-of-the-art algorithms available in the literature on the Boston housing dataset to evaluate the effectiveness of the proposed model. The results show that the proposed model provides better accuracy and outperforms existing algorithms in different evaluation metrics. The code for the implementation is available at   \href{https://github.com/wijayarobert/MultiLevelDenseLayerNN}{https://github.com/wijayarobert/MultiLevelDenseLayerNN}
\end{abstract}

\section{Introduction}
\label{sec:introduction}
Estimating the value of the house is a major problem for many stakeholders. Several factors like
the size of the house, number of rooms, as well as location affect the price of the house. Nevertheless,
the price can be predicted with various different methods. One of the common techniques to use is
regression techniques that involve one or more features as input and single target output. In \cite{a}, the author developed a Support Vector Regression model with Gaussian Filter to predict the housing
price. Other researchers also propose a neural network model with a promising testing accuracy
of the Boston dataset (87.7\%) \cite{b}. This result indicates that a neural network is a viable algorithm for resolving difficult problems, even providing a more robust model than the traditional hedonic model.

The neural network is a widely used machine learning algorithm because of its performance to learn from the raw data. This is because a neural network model can learn automatically the features of the data without requiring extensive handcrafted features beforehand. In a deep architecture of the neural network, multiple layers are involved to represent and perform a non-linear transformation of the data in a different hierarchy. The effectiveness of the neural network algorithm leads the author to develop an appropriate architecture to forecast the housing price with better accuracy.

In this paper, the author proposes a novel neural network model to improve the performance of housing price prediction. Unlike the regular feedforward neural network where the input moves in only one direction, the author designed a three-level neural network that is capable to process the information simultaneously. The results of the proposed model outperform the existing model with the testing accuracy of the Boston dataset reaching 91.1\% accuracy. To verify the methods, the author visualizes the predicted and actual values through a regression graph and plots a histogram to see the frequency of error prediction.


This paper is structured as follows: First, in section \hyperref[sec:problem_settings]{2} the author briefly states the problem definition, including the dataset, performance criteria, and data preprocessing. Section \hyperref[sec:methodologies]{3} describes the architecture of the proposed Neural Network. Then, section \hyperref[sec:experimental_results]{4} illustrates the performance of both the proposed method and existing algorithms. Section \hyperref[sec:conclusion]{5} summarizes the conclusion of this research.

\section{Problem Settings}
\label{sec:problem_settings}
The prediction results of the proposed neural network in \cite{b} are robust and promising. However, all the evaluated models and the final proposed model is a standard feedforward neural networks without any fine-tuning and customization in the development of the model. Despite the higher accuracy achieved in the research, the model still can be improved by applying some advanced features and empirical design of neural network architecture.

\textbf{Dataset:} The Boston Housing Dataset is chosen in this work. The dataset is considered small but widely used, containing 506 cases of housing price information in different suburbs of Boston, Massachusetts. From the original dataset, 405 samples are treated as training data and 101 samples are treated as test data. The Boston dataset covers 14 attributes for each case, an explanation of 14 attributes is shown in Table \ref{tab:table1}.

\begin{table}[h!]
  \begin{center}
    \caption{Attributes explanation of the Boston dataset.}
    \label{tab:table1}
    \begin{tabular}{c|l}
      \textbf{Attributes} & \textbf{Description} \\
      \hline
      CRIM & per capita crime rate by town  \\ 
      ZN & the proportion of residential land zoned for lots over 25,000 sq. ft. \\ 
      INDUS & the proportion of non-retail business
acres per town \\ 
      CHAS & Charles River dummy variable (= 1 if
tract bounds river;
= 0 otherwise) \\
      NOX & nitric oxides concentration (parts
per 10 million) \\
      RM & the average number of rooms per
dwelling  \\
      AGE & the proportion of owner-occupied
units built before 1940 \\
      DIS & weighted distances to five Boston
employment centres \\
      RAD & index of accessibility to radial
highways \\
      TAX & full-value property-tax rate per
\$10,000  \\
      PTRATIO & pupil-teacher ratio by town\\
      B & 1000 (Bk-0.63)2 where Bk is the
proportion of blacks by town \\
      LSTAT & \% lower status of the population\\
      MEDV & The median value of owner-
occupied homes is \$1000's \\
    \end{tabular}
  \end{center}
\end{table}

\textbf{Evaluation metrics:} Several metrics are used to evaluate the accuracy of the proposed model. This includes R-Squared, Mean Absolute Error (MAE), Mean Squared Error (MSE), and Root Mean Squared Error (RMSE). In a regression problem, the coefficient of determination (R$^2$) is commonly used to measure the correlation between the actual and predicted outputs. MAE, MSE, and RMSE are used to assess the error of the model. Specifically, MAE evaluates the absolute distance between the actual and predicted values, while MSE represents the average of the squared difference between true and predicted values. RMSE will take the square root of MSE, and keep maintaining the property of the errors.  The calculation of the metrics is based on the following equations: 
\begin{equation}
    R^2 = 1 - \frac{\sum(y_i - \hat{y})^2}{\sum(y_i - \bar{y})^2}
\end{equation}
\begin{equation}
    MAE = \frac{1}{N} \sum_{i=1}^N |y_i - \hat{y}|
\end{equation}
\begin{equation}
    MSE = \frac{1}{N} \sum_{i=1}^N (y_i - \hat{y})^2
\end{equation}
\begin{equation}
    RMSE = \sqrt{MSE} = \sqrt{\frac{1}{N} \sum_{i=1}^N (y_i - \hat{y})^2}
\end{equation}
\begin{center}
    where $\hat{y}$ is predicted value of $y$, and $\bar{y}$ is the mean value of $y$. 
\end{center}
In general, the R$^2$ represents the accuracy of the model, the higher score of R$^2$, the better model fits, and the lower score of MAE, MSE, and RMSE indicate the better accuracy of the model.

\textbf{Data Normalization:} Before training the model, both the training and testing dataset is normalized. This paper performed standard normalization or z-score normalization by the TensorFlow inbuilt function which transforms the features to have a mean of 0 and a standard deviation of 1. The following equation describes how data normalization is implemented:
\begin{equation}
    z = \frac{x - \mu}{\sigma}
\end{equation}
\begin{center}
    where $\mu = \frac{1}{n}\sum_{i-1}^n x_i$, and  $\sigma = \sqrt{\frac{1}{n} \sum_{i=1}^n (x_i-\mu)^2}$
\end{center}

\section{Methodologies}
\label{sec:methodologies}

\begin{figure}[h]
    \centering
    \includegraphics[width=12cm]{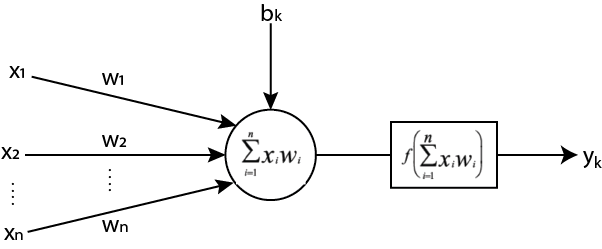}
    \caption{The ANN receive input features $x_1, x_2, ..., x_n$ and multiplied them by weight $w_1, w_2, ..., w_n$. The weighted inputs are summed together with the bias $b$ value before passing through the activation function to produce the net input of the neuron.}
    \label{fig:nn}
\end{figure}

Artificial neural network (ANN) algorithms are inspired by the architecture of neurons in the brain. Through the dynamics of the network, the algorithm can learn to recognize the pattern in a dataset and generalize what they have processed. Given a training set contains a list of input with corresponding labels, the ANN can be trained to classify the data and make some adjustments based on the value between their neurons \cite{c}. This is called supervised learning, where the neuron in the network tries to fit an input-output function with respect to the training data. This understanding can be translated into a mathematical model as shown below:
\begin{equation}
    z = \sum_{i=1}^n x_i w_i + b
\end{equation}
Each input feature $x_1, x_2, ..., x_n$ is multiplied by a weight $w_1, w_2, ..., w_n$ before they are summed together. Then, the constant value called bias ($b$) is added to produce the net input ($z$) of the neuron. The net input will be passed through an activation function ($g$) to produce the new output ($j$) which will be passed on to other neurons. 
\begin{equation}
    j = g(z) = g(\sum_{i=1}^n x_i w_i + b)
\end{equation}
The neuron is responsible to receive information from other neurons, processing the information, and transmitting the result to other neurons. This process is illustrated in Figure \ref{fig:nn}.

\subsection{Architecture}
\begin{figure}[h]
    \centering
    \includegraphics[width=12cm]{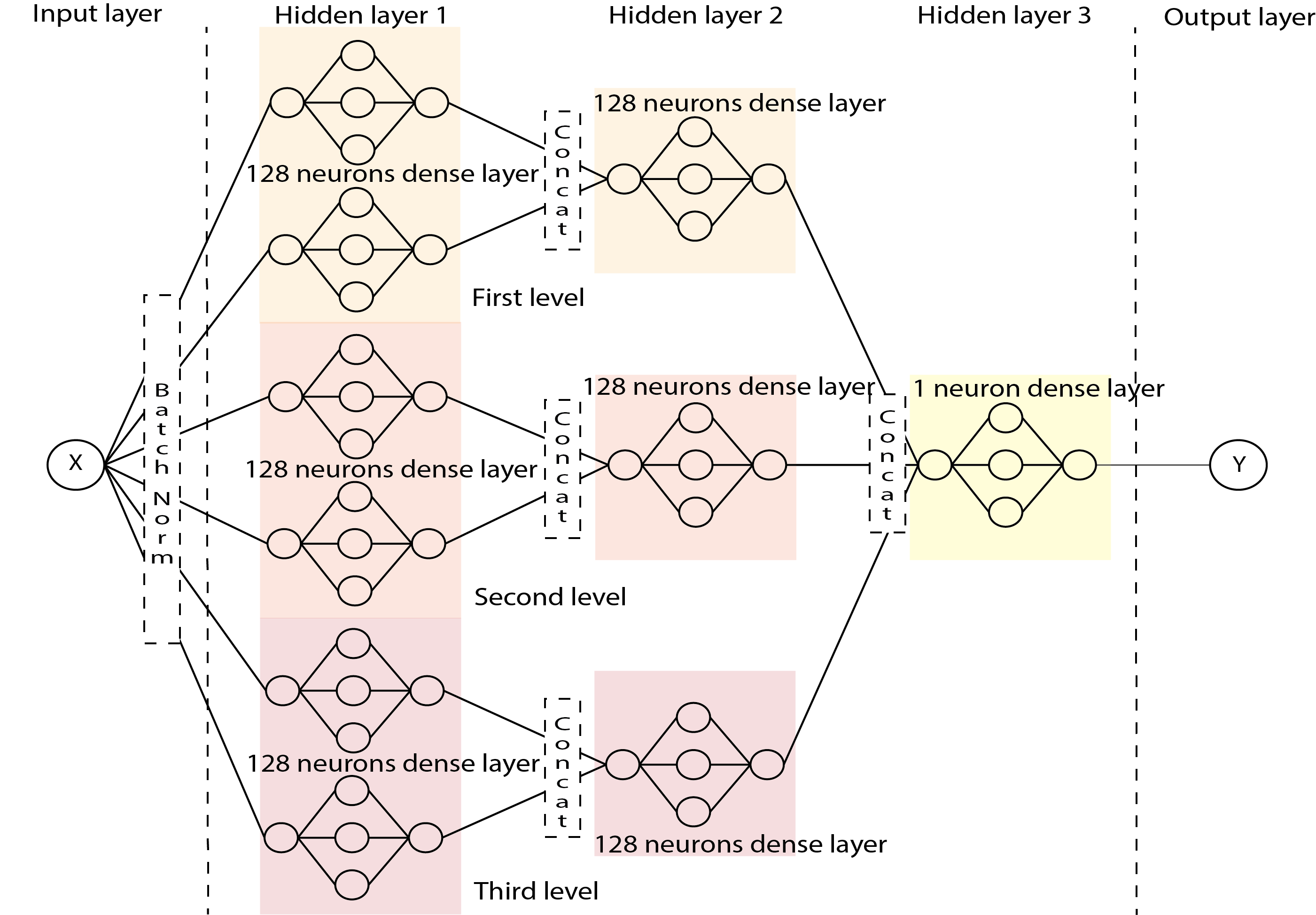}
    \caption{The architecture of the multi-level dense layer neural network. This architecture consists of 10 dense networks in total, spreading in three hidden layers. The batch normalization function is also applied to input data before passing into the hidden layers. The output from the hidden layers will be passed into the output layer as a single unit (the price prediction number).}
    \label{fig:architecture}
\end{figure}

Inspired by the modular neural network where the network consists of several multilayer perceptrons, in this paper the author proposed a multi-level dense layer neural network to yield better generalization. Each module or level in the networks is independent which allows the system to work in parallel \cite{d}. Moreover, from a computational perspective, the modular design of the neural network leads the model to more robust and efficient computation because it eliminates a high coupling burden that is often encountered in a standard/monolithic neural network.\newline\newline
The architecture is shown in Figure \ref{fig:architecture}. As shown in this figure, the network has 3 hidden layers consisting of 10 dense layers in total with the same amount of neurons (128 neurons) that implement the ReLU activation function. The first layer is the input layer of 13 units since the data have 13 features. This input layer will be passed to six (3 pairs) dense layers in the first hidden layer. Each pair have the same amount of neurons, this is because based on experiments conducted by the author, it will lead to better accuracy compared to using different amount of neurons. The output from each level in the first layers will be concatenated as an input for the second hidden layer. The results of three dense layers in the second hidden layer will be combined together before being calculated in the last hidden layers by a single dense layer. The output from this process will be passed into the output layer with a single unit, which is the price prediction of the house.

\begin{figure}[h]
    \centering
    \includegraphics[width=12cm]{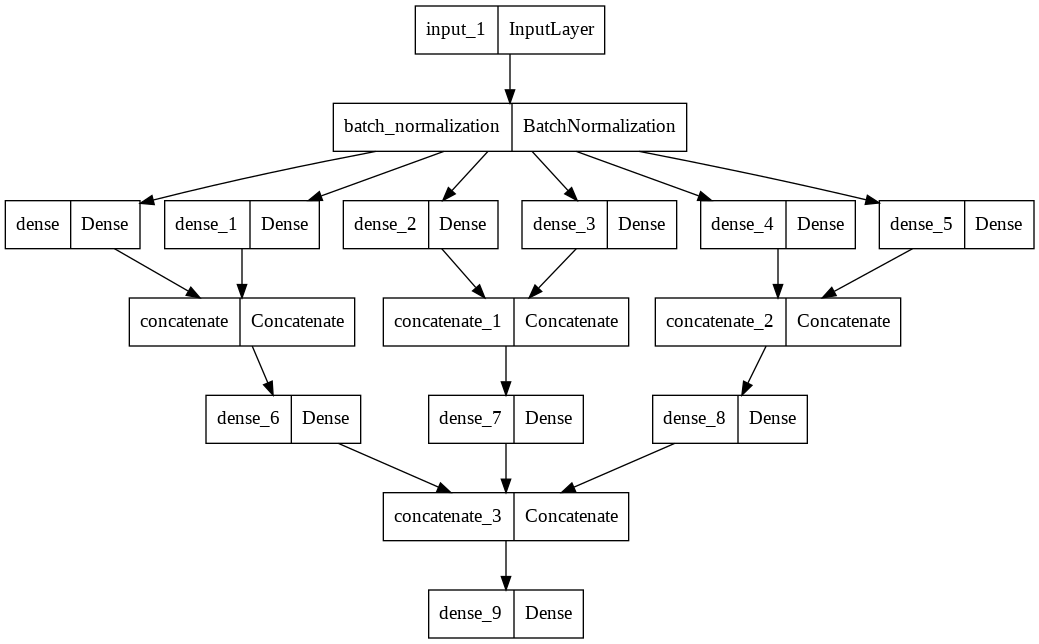}
    \caption{Network architecture generated with Keras.util plot\_model function. In summary, the network consists of 1 input layer, 1 batch normalization function, 10 dense layers, 4 concatenating processes, and 1 output layer.}
    \label{fig:architecture_tensorflow}
\end{figure}

\subsection{Batch Normalization}
The standard Batch Normalization (BN) was implemented in the early part of the model architecture to accelerate the training process of the network. Although the precise effect of BN remains a topic of further investigation \cite{e}, the experiments during model development in this paper show that BN has positive effects on several aspects of neural networks. According to \cite{e}, BN helps reduce the internal covariate shift caused by the change of distribution of the input signal. In this experiment, it helps speed up the training process and benefits every measurement metric used in this research. 



\subsection{Training Process}
\begin{figure}[h]
\par
\raisebox{-.5\height}{\includegraphics[width=7cm]{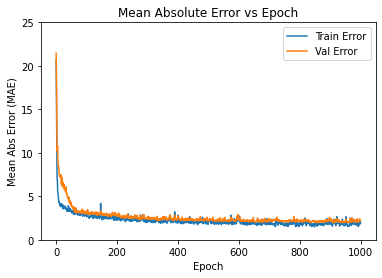}}%
\hfill
\raisebox{-.5\height}{\includegraphics[width=7.2cm]{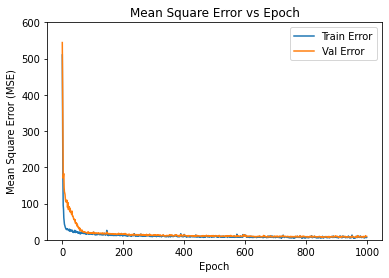}}%
\par
\caption{MAE and MSE score during training of the model. Both MAE and MSE decreased significantly in the first 50 epochs of the training. For additional information, it takes 1 minute 44 seconds, and 1 minute 25 seconds for training the model using CPU (AMD Ryzen 3), and Google Colab GPU, respectively.}
\label{fig:MAE_MSE}
\end{figure}

The developed model was trained and validated by 20\% of the training dataset for 1000 epochs. In the training of the model, the author utilizes the Adam (Adaptive Moments) optimization method with a learning rate equal to 0.001. The optimizer is responsible to optimize the loss function during training and adjust the attributes of the neural network. This will helps in reducing the loss and increase the accuracy. The MAE and MSE during training are shown in Figure \ref{fig:MAE_MSE}. 

\section{Experimental Results}
\label{sec:experimental_results}

In this section, the author has performed experiments to measure the accuracy of the model and compare the proposed model with existing algorithms in \cite{b}.  

\subsection{Results on Boston Dataset}

\begin{figure}[h]
\par
\raisebox{-.5\height}{\includegraphics[width=5cm]{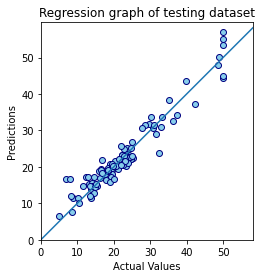}}%
\hfill
\raisebox{-.5\height}{\includegraphics[width=7.6cm]{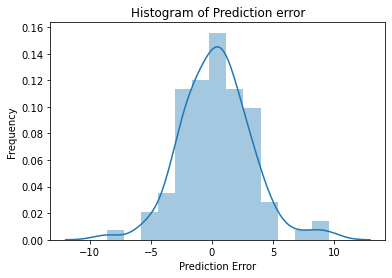}}%
\par
\caption{Regression graph and histogram of prediction error on the testing dataset.}
\label{fig:Regression_Histogram}
\end{figure}

To verify the accuracy during training, the model was evaluated with the testing dataset. The regression graph of the testing dataset and histogram of prediction is shown in  Figure \ref{fig:Regression_Histogram}. As shown in the figure, the regression graph shows that the proposed neural network model is capable to construct a correlation between input and output parameters and give robust prediction results. The values of the four evaluation metrics for both the training and testing dataset are shown in Table \ref{tab:table2}. The R$^2$ value is 0.948, and 0.911 for the training and testing dataset, respectively. The MAE value for the training dataset is 1.99, and 2.31 for the testing dataset. The MSE value is 7.24 and 9.16, respectively for the training and testing dataset. Lastly, the RMSE value corresponding to the training set is 2.69, and 3.02 for the testing set. These values indicate that the model is robust and provides better accuracy compare to the current existing algorithm. Furthermore, to see the performance of the model, the author compares the actual value (the price of the house) with the predicted value generated by the proposed model as shown in table \ref{tab:table3}.
 
\begin{table}[h!]
  \begin{center}
    \caption{True value versus predicted value generated by the proposed model.}
    \label{tab:table3}
    \begin{tabular}{c|l}
      \textbf{True Value} & \textbf{Predicted Value} \\
      \hline
      18.9 & 18.396809  \\ 
      13.9 & 14.814673 \\ 
      18.4 & 18.561754 \\ 
      14.5 & 17.794323 \\
      13.2 & 10.939665 \\
      24.7 & 23.025442  \\
      30.8 & 31.412012 \\
      25.3 & 26.502363 \\
      16.6 & 17.418531 \\
      23.4 & 23.580996  \\
    \end{tabular}
  \end{center}
\end{table}

\subsection{Algorithm Comparison}
\begin{figure}[h]
    \centering
    \includegraphics[width=12cm]{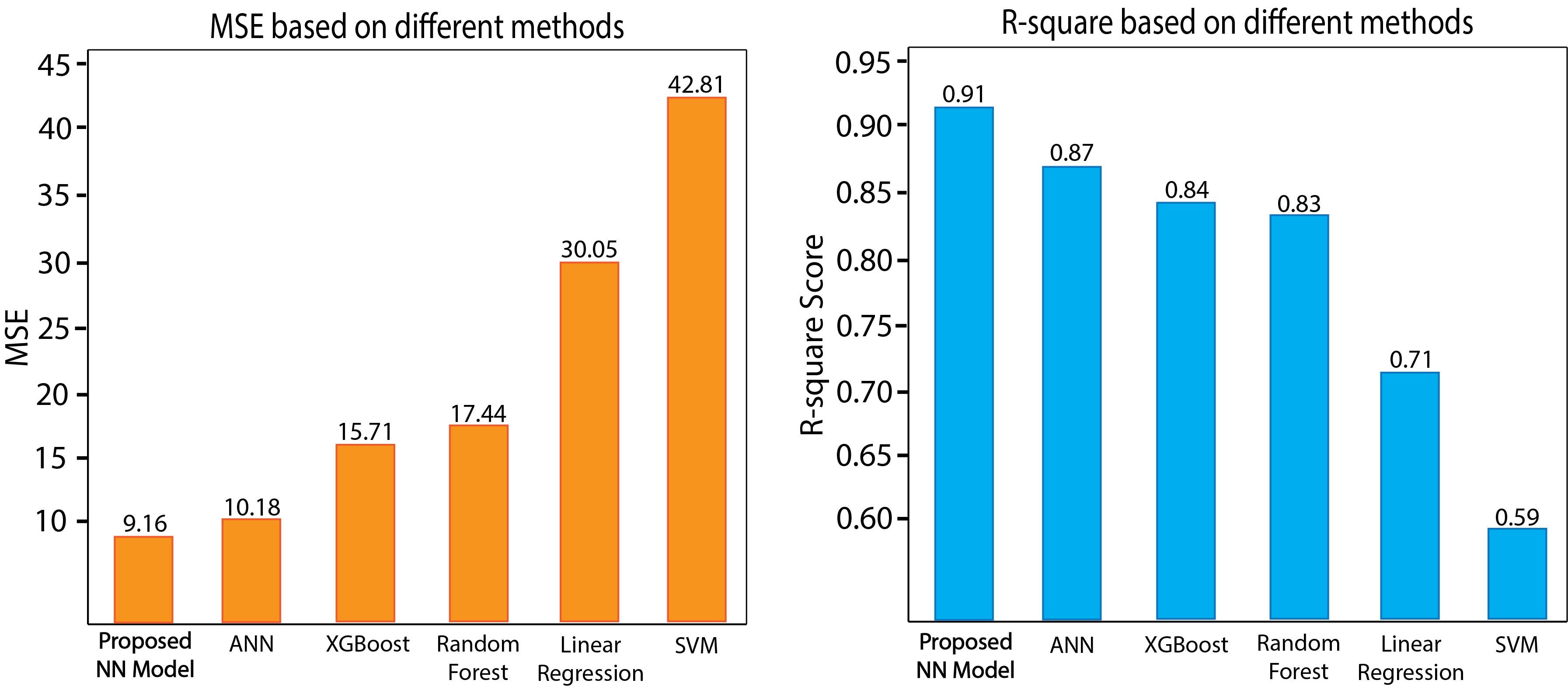}
    \caption{Comparison of R$^2$ and MSE score on different algorithms. As shown in the figure, the proposed NN model outperforms existing algorithms based on R$^2$ and MSE metrics.}
    \label{fig:MSE_R2}
\end{figure}

Several existing algorithms available in the literature are compared with the proposed model. Figure \ref{fig:MSE_R2} and table \ref{tab:table4} described the performance of the proposed model compare to the existing algorithms. As it shown in the figure \ref{fig:MSE_R2} and table \ref{tab:table4}, the proposed model provides more robust results in most evaluation metrics.


\begin{table}[h!]
  \begin{center}
    \caption{Algorithm comparison on different evaluation metrics. The proposed model outperforms existing algorithms in three different metrics (R$^2$, MSE, and RMSE)}
    \label{tab:table4}
    \begin{tabular}{l|c|c|c|c}
      \textbf{Algorithms} & 
      \textbf{R$^2$} &
      \textbf{MSE} & 
      \textbf{RMSE} & 
      \textbf{MAE} \\
      \hline
      \textbf{Proposed NN} & \textbf{0.91} & \textbf{9.16} & \textbf{3.02} & 2.31\\ 
      ANN \cite{b} & 0.87 & 10.18 & 3.19 & 2.10\\ 
      XGBoost \cite{b} & 0.84 & 15.71 & 2.45 & 2.45\\
      Random Forest \cite{b} & 0.83 & 17.44 & 4.17 & 2.56\\
      Linear Regression \cite{b} & 0.71 & 30.05 & 5.48 & 3.85\\
      SVM \cite{b} & 0.59 & 42.81 & 6.54 & 3.75\\
    \end{tabular}
  \end{center}
\end{table}
\newpage

\section{Conclusion}
\label{sec:conclusion}
In this paper, the author developed a novel neural network model to improve the performance of the housing price prediction model using the Boston housing dataset. The architecture of the model is inspired by a modular neural network where the network consists of several independent multilayer perceptrons that work in parallel. The proposed model is evaluated by examining the performance based on different metrics, namely R$^2$, MAE, MSE, and RMSE. The experimental result shows that the proposed model significantly outperforms the existing state-of-the-art algorithms including ANN, XGBoost, Random Forest, Linear Regression, and SVM that have been utilized in the literature. Admittedly, the prediction accuracy is still limited in the Boston dataset, and the universality of the model needs to be measured in further research. As in the further study, the author would like to further explore novel neural network techniques and architecture that can be applied to a wider variety of real-world problems in society.

\small{



}

\end{document}